\documentclass[prb,reprint]{revtex4-1}

\usepackage{graphicx}
\usepackage{dcolumn}
\usepackage{bm}

\begin{document}

\title{Manipulation of the Land$\acute{\text{e}}$ g-factor  in InAs quantum dots through the
application of anisotropic gate potentials: Exact diagonalization, numerical and perturbation methods}

\author{Sanjay Prabhakar,$^{1}$ James E. Raynolds$^{2}$ and  Roderick Melnik$^{1,3}$}
\affiliation{
$^1$M\,$^2$NeT Laboratory, Wilfrid Laurier University, Waterloo, ON, N2L 3C5 Canada\\
$^2$Sterne, Kessler, Goldstein $\&$ Fox P.L.L.C.,1100 New York Avenue, NW, Washington, DC 20005\\
$^3$Department of Mathematical Information Technology, University of Jyv$\mathrm{\ddot{a}}$skyl$\mathrm{\ddot{a}}$, 40014, Finland}

\date{September 27, 2011}

\begin{abstract}
We study the variation in the Land$\acute{\text{e}}$ g-factor of electron spins induced by both anisotropic gate potentials and magnetic fields  in InAs quantum dots for  possible implementation towards solid state quantum computing.  In this paper, we present  analytical expressions and numerical
simulations of the variation in the Land$\acute{\text{e}}$ g-factor for both isotropic and anisotropic quantum dots.
Using both  analytical techniques and numerical simulations,
we show that the Rashba spin-orbit coupling has a major contribution in the variation of the g-factor with electric fields  before the regime, where  level crossing or anticrossing occurs. In particular, the electric field tunability is shown to cover a wide range of g-factor through strong Rashba spin-orbit interaction.
Another major result
of this paper is that the anisotropic gate potential gives  quenching effect in
the orbital angular momentum that reduces the variation in
the E-field  and B-field tunability of the g-factor if  the area of the symmetric and asymmetric quantum dots is held constant.
We  identify  level crossings and  anticrossings  of the electron states in the variation of the Land$\acute{\text{e}}$ g-factor. We  model the wavefunctions of electron spins and estimate the size of the anticrossing for the spin states  $|0,-1,+1/2>$ and $|0,0,-1/2>$ corresponding to a quantum dot that has been recently studied experimentally (Phys. Rev. Lett. \textbf{104}, 246801 (2010)).

\end{abstract}

\maketitle

\section{Introduction}
Single electron spins in an electrostatically defined quantum dot in a 2-dimensional electron gas (2DEG) have been manipulated and studied by several groups.\cite{bandyopadhyay00,awschalom02,prabhakar09,prabhakar10,Pryor06,brum97,sousa03}  Quantum dots in III-V type semiconductors  provide an  opportunity to study the variation in the Land$\acute{\text{e}}$ g-factor vs. gate potentials and magnetic fields.\cite{engel04,levitov03,wang10,chang03,rahman09,de09} The shape and size of the quantum dots can be modified by changing the gate controlled electric fields that influence   the variation in the energy spectrum as well as  the Land$\acute{\text{e}}$ g-factor of the dots.\cite{prabhakar09,nowack07,nakaoka07}  The results of this research  might enhance the opportunities  of building spintronic logic devices for  possible implementation towards solid state quantum computing.\cite{rashba03a,loss98,rashba05,zutic04,shim08}

The orbital and spin angular momentum of the electron  in a semiconductor quantum dot   interact through the Rashba and Dresselhaus spin-orbit couplings.~\cite{bychkov84,dresselhaus55}
These two spin-orbit coupling effects arise from two different
types of symmetry operations in III-V type semiconductors.
The Rashba spin-orbit coupling arises from the structural inversion asymmetry
of the triangular shaped quantum well confining potential. The Dresselhaus spin-orbit coupling arises from bulk inversion asymmetry.
The mathematical
expressions for the Rashba and Dresselhaus spin-orbit couplings that are implemented into the theoretical model  are given in this paper in Section II. The strength of the Rashba and Dresselhaus spin-orbit couplings is determined by the gate controlled electric fields  and  is an   important parameter in controlling the Land$\acute{\text{e}}$ g-factor for both isotropic and anisotropic quantum dots.

The goal of the present work is to explore the non-degenerate  energy spectrum of electrostatically defined  InAs quantum dot by both analytical techniques and numerical simulations. These approaches provide  realistic information for controlling the Land$\acute{\text{e}}$ g-factor for both isotropic and anisotropic quantum dots through the application of gate potentials. We also model  realistic wavefunctions of electrons in InAs quantum dots that were recently studied by  experimentalists in Ref. \onlinecite{takahashi10}. We estimate   the size of avoided anticrossing is approximately $65$ $\mu$eV, which is in    agrement with the experimentally  reported values.~\cite{takahashi10}
In this paper, by utilizing both analytical and numerical techniques, we find that the  Rashba spin-orbit coupling produces the dominant effect on the variation of the Land$\acute{\text{e}}$ g-factor vs. electric field strength below the level crossing or anticrossing. Also, anisotropic gate potential lead to a quenching effect in the orbital angular momentum that reduces the variation in the g-factor. Our work is similar to those of Refs.~\onlinecite{prabhakar09},~\onlinecite{prabhakar10},~\onlinecite{Pryor06},~\onlinecite{pingenot08},~\onlinecite{destefani05} and \onlinecite{olendski07} but differs in  that we utilize both analytical and numerical approaches based on the finite element method.

The paper is organized as follows. In Sec. II, we exactly diagonalize the Hamiltonian of a quantum dot confined in an asymmetric potential  including spin-orbit interaction and  a magnetic field along z-direction.  In Sec. III, based on a second order perturbation calculation, we present  analytical expressions of the Land$\acute{\text{e}}$ g-factor for both isotropic and anisotropic quantum dots. In Sec. IV, we plot the Land$\acute{\text{e}}$ g-factor induced by gate controlled electric fields  vs. magnetic field as well as quantum dot radii for pure Rashba case ($\alpha_D=0$), pure Dresselhaus case ($\alpha_R=0$) and mixed cases (both $\alpha_R,\alpha_D$ present) where $\alpha_R$ and $\alpha_D$ are two parameters related to the strength of the Rashba and Dresselhaus spin-orbit interactions respectively. We also model the wavefunctions of electron states that were recently reported by experimentalists in Ref.~\onlinecite{takahashi10}. Finally, in Sec. V, we summarize our results.
\begin{figure}
\includegraphics[width=9cm,height=6cm]{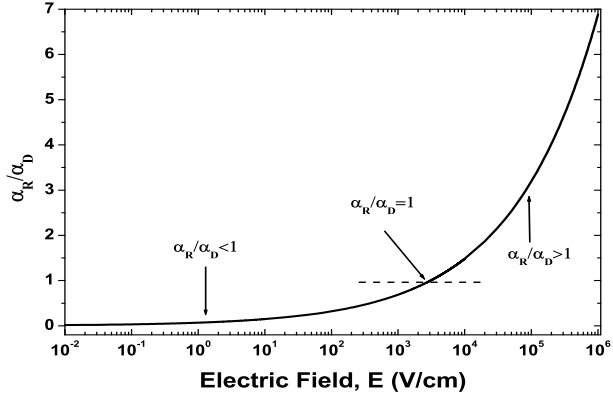}
\caption{\label{fig1} Contributions  of Rashba and Dresselhaus spin-orbit couplings ($\alpha_R/\alpha_D$) vs. Electric Field.  It can be seen that Rashba and Dresselhaus spin-orbit couplings becomes equal at the electric field, $E=3.05\times10^3$ V/cm. This plot is obtained from Eq.~\ref{coefficient-R-D}.}
\end{figure}

\section{Theoretical Model}
The  Hamiltonian of an electron in a quantum dot in the plane of a 2DEG, in the presence of an external magnetic field along z-direction can be written as~\cite{sousa03,khaetskii00}
\begin{equation}
H = H_{xy} +  H_{so},
\label{total}
\end{equation}
where the Hamiltonian $H_{so}$ is associated to  the Rashba-Dresselhaus spin-orbit couplings and $H_{xy}$ is the Hamiltonian of the electron along the lateral direction in the plane of the 2DEG. $H_{xy}$ can be written as
\begin{equation}
H_{xy} = {\frac {\vec{P}^2}{2m}} + {\frac{1}{2}} m \omega_o^2
(a x^2 + b y^2) + {\frac 1 2} g_o \mu_B \sigma_z B,
\label{hxy}
\end{equation}
where $\vec{P} = \vec{p} + e \vec{A}$ is  the kinetic momentum operator, $\vec{p} = -i\hbar (\partial_x,\partial_y,0)$ is canonical momentum operator and $\vec{A} = {\frac {B}{a+b}} (-yb,xa,0)$ is the vector potential in the asymmetric gauge. Here, $-e < 0$ is the electronic charge, $m$ is the effective mass of the electron in the conduction band,  $\mu_B$ is the Bohr magneton, $\sigma_z$ is the Pauli spin matrix along z-direction.  Also, $\omega_0=\frac{\hbar}{m\ell_0^2}$ is  a parameter characterizing the strength of the confining potential and $\ell_0$ is used as the radius of the quantum dot. By choosing the appropriate values of  $a$ and $b$, one can define the shape of the quantum dot from circle to ellipse. The Hamiltonian $H_{xy}$ can be exactly diagonalized (see Appendix A) by writing the Hamiltonian as a function of number operators $n_{\pm}=a_{\pm}^\dagger a_{\pm}$ and its energy spectrum can be written as,~\cite{schuh85}
\begin{widetext}
\begin{eqnarray}
\varepsilon_{n_+n_-}=\left(n_++n_-+1\right)\hbar\omega_++\left(n_+-n_-\right)\hbar\omega_-+ {\frac 1 2} g_o \mu_B \sigma_z B, \label{epsilon}\\
a_{\pm}=\frac{1}{\left(s_+-s_-\right)\left(1+i\right)}
\left[
\pm\left(s_{\mp}\pm i\right)\frac{\ell}{\hbar} p_x+\left(s_{\pm}\pm i\right)\frac{\ell}{\hbar} p_y
+\left(1\mp is_{\mp}\right)\frac{1}{\ell} x \pm \left(1\mp is_{\pm}\right)\frac{1}{\ell} y \right],\\ \label{a-lowering}
a_{\pm}^\dagger=\frac{1}{\left(s_+-s_-\right)\left(1-i\right)}
\left[
\pm\left(s_{\mp}\mp i\right)\frac{\ell}{\hbar} p_x+\left(s_{\pm}\mp i\right)\frac{\ell}{\hbar} p_y
+\left(1\pm is_{\mp}\right)\frac{1}{\ell} x \pm \left(1\pm is_{\pm}\right)\frac{1}{\ell} y \right],\label{a-raising}
\end{eqnarray}
\end{widetext}
where,
\begin{eqnarray}
s_{\pm}=\frac{\omega_+}{\omega_c\sqrt{\frac{\Omega_2}{\Omega_1}}}\left[\frac{\Omega_2}{\Omega_1}-1 \pm \sqrt{\frac{\omega_c^2\frac{\Omega_2}{\Omega_1}}{\omega_+^2}+\left(1-\frac{\Omega_2}{\Omega_1}\right)^2}\right],\label{s-pm}~~~\\
\omega_{\pm}=\frac{1}{2}\left[\omega_c^2+\left(\Omega_1\pm\Omega_2\right)^2\right]^{1/2},~~~~\label{omega-pm}
\end{eqnarray}
provided that $\left[a_{\pm},a_{\pm}^\dagger\right]=1$ and  $\left[x,p_x\right]=\left[y,p_y\right]=i\hbar$. Here $\Omega_1=\omega_0\sqrt{a}$, $\Omega_2=\omega_0\sqrt{b}$, $\ell=\sqrt{\frac{\hbar}{m\Omega}}$, $\Omega=\sqrt{\omega_0^2+\frac{1}{4}\omega_c^2}$ and $\omega_c=\frac{eB}{m}$ is the cyclotron frequency. Note that a similar type of expression for the energy spectrum of an anisotropic quantum dot is also discussed in Ref. \onlinecite{olendski07}. However, our methodology for diagonalizing the Hamiltonian is slightly different (see in Appendix A).
Also,  we verified that by substituting $a=b=1$,  Eqs. (\ref{epsilon}-\ref{s-pm}) are exactly the same as those in  Ref.~\onlinecite{sousa03}.
The eigenstates of $H_{xy}$ in Eq.~\ref{hxy} with $a=b=1$ are
well-known Fock-Darwin energy states.~\cite{fock28,darwin30}

The Hamiltonian associated with the spin-orbit couplings  can be written as~\cite{bychkov84,dresselhaus55}
\begin{equation}
H_{so} =\frac{\alpha_R}{\hbar}\left(\sigma_x P_y - \sigma_y P_x\right)+ \frac{\alpha_D}{\hbar}\left(-\sigma_x P_x + \sigma_y P_y\right).\label{rashba-dresselhaus}
\end{equation}
The spin-orbit Hamiltonian consists of the Rashba coupling whose strength is characterized
by the parameter $\alpha_R$ and the  Dresselhaus coupling with $\alpha_D$. These coupling parameters depend on the
electric field E of the quantum well confining potential (i.e., $E=-\partial V/\partial z$) along z direction at the interface in a heterojunction
as
\begin{equation}
\alpha_R=\gamma_ReE,~~~~~\alpha_D=0.78\gamma_D\left(\frac{2me}{\hbar^2}\right)^{2/3}E^{2/3},\label{coefficient-R-D}
\end{equation}
where the Rashba coefficient $\gamma_R=110 {\AA}^2$, Dresselhaus coefficient $\gamma_D=130~eV{\AA}^3$ and effective mass $m=0.0239$ have been considered  for InAs quantum dots.~\cite{sousa03}

The Rashba and  Dresselhaus spin-orbit Hamiltonian can be written in terms of raising and lowering operators  as:~\cite{sousa03}
\begin{widetext}
\begin{eqnarray}
H_{so}&=&\alpha_R \left(1+i\right)\left[
b^{1/4}\kappa_+\left(s_+-i\right)a_++b^{1/4}\kappa_+\left(s_-+i\right)a_-+a^{1/4}\eta_-\left(i-s_-\right)a_++a^{1/4}\eta_-\left(i+s_+\right)a_-
\right]\nonumber\\
&&+\alpha_D
\left(1+i\right)\left[a^{1/4}\kappa_-\left(i-s_-\right)a_++a^{1/4}\kappa_-\left(i+s_+\right)a_-+b^{1/4}\eta_+\left(-i+s_+\right)a_++b^{1/4}\eta_+\left(i+s_-\right)a_-\right]+h.c.,~~~~~~
\label{H-R}
\end{eqnarray}
\end{widetext}
where,
\begin{eqnarray}
\kappa_{\pm}=\frac{1}{2\left(s_+-s_-\right)}\left\{\frac{1}{\ell}\sigma_x\pm i\frac{eB\ell}{\hbar}\left(\frac{1}{\sqrt{a}+\sqrt{b}}\right)\sigma_y\right\},\nonumber\\
\eta_{\pm}=\frac{1}{2\left(s_+-s_-\right)}\left\{\frac{1}{\ell}\sigma_y\pm i\frac{eB\ell}{\hbar}\left(\frac{1}{\sqrt{a}+\sqrt{b}}\right)\sigma_x\right\},\nonumber
\end{eqnarray}
and h.c. represents the hermitian conjugate. It is clear that the spin-orbit Hamiltonian and the Zeeman spin splitting energy in both isotropic and anisotropic quantum dots obeys a selection rule in which the orbital angular momentum can change by one quantum.

\section{Analytical expression for the Land$\acute{\text{E}}$ $g$-factor }
Our conventional definition of the   electron g-factor in the conduction band in the presence of magnetic field along z-direction can be written as~\cite{Pryor06,sousa03}
\begin{equation}
g = {\frac {\varepsilon_{0,0,+1/2} - \varepsilon_{0,0,-1/2}} {\mu_B B}}, \label{g}
\end{equation}
where $\varepsilon_{0,0,+1/2}$ and  $\varepsilon_{0,0,-1/2}$ are the eigenvalues of the electron in the conduction band  with spin up and down respectively having the lowest orbital angular momentum.

At low electric fields and small quantum dot radii, we treat the Hamiltonian associated to the Rashba and Dresselhaus spin-orbit couplings as a perturbation. Based on second order perturbation theory, the energy of the electron spin states can be written as,
\begin{eqnarray}
\varepsilon_{0,0,+1/2}=\varepsilon^{(0)}_{0,0,+1/2}+\varepsilon^{(2)}_{0,0,+1/2},\label{epsilon-b}\\
\varepsilon_{0,0,-1/2}=\varepsilon^{(0)}_{0,0,-1/2}+\varepsilon^{(2)}_{0,0,-1/2}.\label{epsilon-c}
\end{eqnarray}
The zero order energy correction can be easily calculated from Eq.~(\ref{epsilon}). The first order energy correction is zero. The calculation for the second order energy correction can be written   as:
\begin{eqnarray}
\varepsilon^{(2)}_{0,0,+1/2}=\frac{\alpha_R^2\xi_++\alpha_D^2\varsigma_+}{\varepsilon^{(0)}_{0,0,+1/2} -\varepsilon^{(0)}_{1,0,-1/2}}+\frac{\alpha_R^2\varsigma_-+\alpha_D^2\xi_-}{\varepsilon^{(0)}_{0,0,+1/2}-\varepsilon^{(0)}_{0,1,-1/2}},~~~~~\\
~\nonumber\\
\varepsilon^{(2)}_{0,0,-1/2}=\frac{\alpha_R^2\varsigma_++\alpha_D^2\xi_+}{\varepsilon^{(0)}_{0,0,-1/2} -\varepsilon^{(0)}_{1,0,1/2}}+\frac{\alpha_R^2\xi_-+\alpha_D^2\varsigma_-}{\varepsilon^{(0)}_{0,0,-1/2}-\varepsilon^{(0)}_{0,1,+1/2}},~~~~~
\end{eqnarray}
where,
\begin{eqnarray}
\xi_{\pm}=\frac{1}{2(s_+-s_-)}\left\{\pm\frac{1}{s_{\pm}}\alpha^2_{\pm}+2\alpha_{\pm}\beta_{\pm} \mp\frac{1}{s_{\mp}}\beta^2_{\pm}\right\},\\
~\nonumber\\
\varsigma_{\pm}=\frac{1}{2(s_+-s_-)}\left\{\pm\frac{1}{s_{\pm}}\alpha^2_{\mp}-2\alpha_{\mp}\beta_{\mp} \mp\frac{1}{s_{\mp}}\beta^2_{\mp}\right\},\\
\alpha_{\pm}=a^{1/4}\left\{\frac{1}{\ell}\pm \frac{eB\ell}{\hbar}\frac{1}{\left(\sqrt{a}+\sqrt{b}\right)}\right\},\\
\beta_{\pm}=b^{1/4}\left\{\frac{1}{\ell}\pm \frac{eB\ell}{\hbar}\frac{1}{\left(\sqrt{a}+\sqrt{b}\right)}\right\}.~
\end{eqnarray}
By substituting Eqs.~\ref{epsilon-b} and~\ref{epsilon-c} in Eq.~\ref{g}, the expression for the
Land$\acute{\text{e}}$ g-factor of anisotropic quantum dots can be written as,
\begin{widetext}
\begin{equation}
g_{\mathrm{asym}}=g_0-\frac{1}{\mu_BB}\left\{
\frac{\alpha_R^2\xi_++\alpha_D^2\varsigma_+}{\hbar\omega_x-\Delta}+\frac{\alpha_R^2\varsigma_-+\alpha_D^2\xi_-}{\hbar\omega_y-\Delta}
-\frac{\alpha_R^2\varsigma_++\alpha_D^2\xi_+}{\hbar\omega_x+\Delta}-\frac{\alpha_R^2\xi_-+\alpha_D^2\varsigma_-}{\hbar\omega_y+\Delta}
\right\}. \label{g-anisotropic}
\end{equation}
\end{widetext}
In the above expression, we use the relation $\omega_x=\omega_++\omega_-$, $\omega_y=\omega_+-\omega_-$ and $\Delta=g_0\mu_BB$. By substituting $a=b=1$ in Eq. \ref{g-anisotropic}, one  find the Land$\acute{\text{e}}$ g-factor for  isotropic quantum dot:
\begin{widetext}
\begin{eqnarray}
g_{sym} & = & g_0 +2 \frac{m_e m}{\hbar^4}\left[\alpha_D^2 \left(1-\delta\right)-\alpha_R^2\left(1+\delta\right)\right]\ell_0^2\nonumber\\
& & -\frac{1}{2}\frac{m_e m^{ 3}}{\hbar^6}\left[\alpha_D^2 \left(\frac{1}{2}-\frac{1}{2}\delta+\delta^2+\delta^3\right)-\alpha_R^2\left(\frac{1}{2}+\frac{1}{2}\delta+\delta^2-\delta^3\right)\right] \omega_c^2\ell_0^6
+\cdots,~
\label{g-isotropic}
\end{eqnarray}
\end{widetext}
where $g_0=-15$ is the bulk g-factor for InAs quantum dots and $\delta=g_0m/m_e$. Here $m_e$ is the mass of electron. The analytical expressions for the Land$\acute{\text{e}}$ g-factor of both anisotropic  and isotropic quantum dots in Eqs.~(\ref{g-anisotropic}) and (\ref{g-isotropic}) respectively are valid before the level crossing or anticrossing occurs.
It can be seen that the g-factor for  both isotropic and anisotropic quantum dots depends on the anisotropic gate potentials, quantum dot radii and magnetic fields. These are our control parameters in manipulating the g-factor  for both isotropic and anisotropic quantum dots.

\begin{figure}
\includegraphics[width=9.5cm,height=10cm]{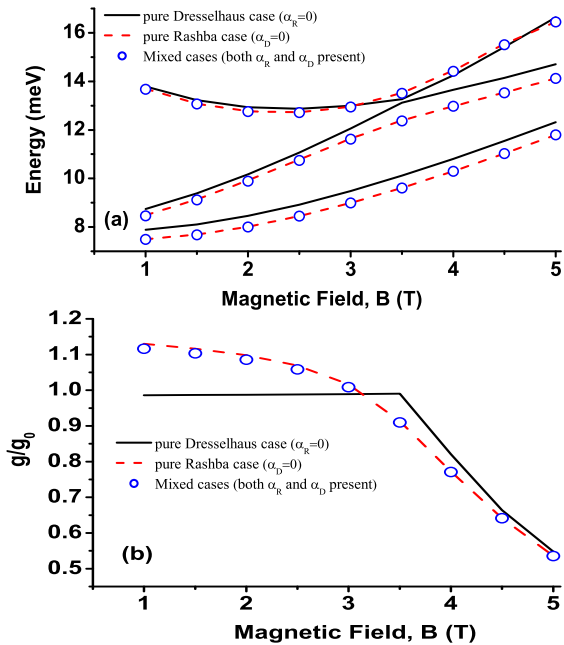}
\caption{\label{fig2} (Color online) (a) Eigenenergy levels for ground, first and second excited states  vs. magnetic fields. (b) Effective Land$\acute{\text{e}}$ g-factor vs. magnetic fields. In both cases, solid lines represent    pure Dresselhaus case ($\alpha_R=0$),  dashed lines represent   pure Rashba case ($\alpha_D=0$) and  open circles represent  mixed cases (both $\alpha_R$ and $\alpha_D$ present). Also in both
cases, we choose the electric field $E=3\times 10^5$ V/cm and the quantum dot radius, $\ell_0=20$ nm  for symmetric  quantum dots ($a=b=1$).}
\end{figure}
\section{Results and Discussions}
We have used the  Finite Element Method\cite{comsol} to solve numerically the corresponding eigenvalue problem with Hamiltonian given by Eq.~(\ref{total})  to  study  the variation in the Land$\acute{\text{e}}$ g-factor at high electric fields for the values $\alpha_R/\alpha_D>1$, large quantum dot radii and magnetic fields for both isotropic ($a=b$) and anisotropic ($a \neq b$)  quantum dots. Throughout the simulations, unless otherwise stated, we consider $\ell_0=20$ nm for $g/g_0$ vs. magnetic field and $B=1$ T for $g/g_0$ vs. quantum dot radius. Our numerical simulations related to the g-factor in quantum dots are valid before the level crossing or anticrossing occurs. At or after the level crossing or anticrossing, g-factor simply captures the effects of two lowest eigenstates in quantum dots.

In Fig.~\ref{fig1}, we investigate the strength of the Rashba and Dresselhaus spin-orbit interactions vs. electric fields.
The strength of the Rashba and Dresselhaus spin-orbit couplings for InAs quantum dots can be determined by the relation $\alpha_R/\alpha_D=0.015E^{1/3}$ (see Eq.~\ref{coefficient-R-D}). From this relation, it can be found that both Rashba and Dresselhaus spin-orbit couplings become equal at the electric field  $E=3.05\times 10^3$ V/cm. However, at this value of the electric field where $\alpha_R=\alpha_D$, the rotational symmetry is not broken and the  spin splitting energy mainly corresponds to the Zeeman energy. For $E=10^4$ V/cm to $10^6$ V/cm, $\alpha_R/\alpha_D$ varies from $1.5$ to $6.88$ (see Fig.~\ref{fig1}). Since $\alpha_R/\alpha_D>1$, only the Rashba spin-orbit coupling has an appreciable contribution to  spin splitting energy.

Figure~\ref{fig2} illustrates the eigenenergy of the states $|0,0,+1/2>$, $|0,0,-1/2>$, and  $|0,-1,+1/2>$  and g-factor vs. magnetic fields of  symmetric quantum dots ($a=b=1$)  for pure Rashba  ($\alpha_D=0$, solid lines), pure Dresselhaus   ($\alpha_R=0$, dashed lines) and mixed cases (both  $\alpha_R$ and $\alpha_D$ present, open circles). It can be seen that the Dresselhaus spin-orbit coupling has almost no effect on the manipulation of the g-factor. For the pure Dresselhaus case, we find a level crossing at $3.5$ T. However, for the pure Rashba and mixed spin-orbit couplings, we find an avoided anticrossings in the manipulation of the g-factor.

\begin{figure}
\includegraphics[width=10cm,height=13cm]{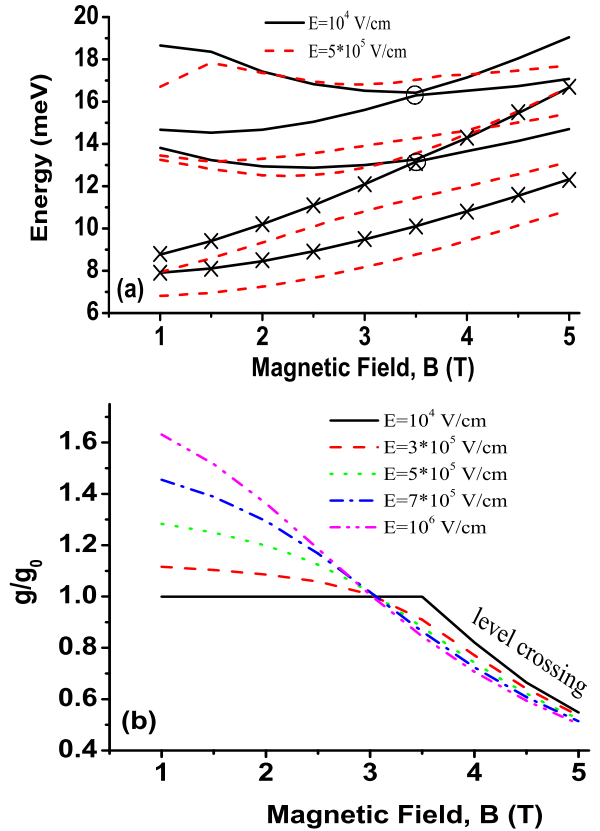}
\caption{\label{fig3} (Color online) (a) Several eigenenergy levels vs. magnetic fields at the electric fields: $E=10^4~V/cm$ (solid lines) and $E=5\times10^5~V/cm$ (dashed lines). The symbol $\times$ represents  data from perturbation theory. A level crossing occurs at the magnetic field,  $B=3.5$ Tesla for the electric field, $E=10^4$ V/cm (solid  line) and is shown by open circles. (b) Effective Land$\acute{\text{e}}$ g-factor vs. magnetic field at the electric fields  $E=10^4$ V/cm (solid line), $3\times10^5$ V/cm (dashed line), $5\times10^5$ V/cm (dotted line), $7\times10^5$ V/cm (dashed-dotted line) and $10^6$ V/cm (dashed-dotted-dotted line). In (a) and (b), we choose the quantum dot radius, $\ell_0=20$ nm for symmetric  quantum dots ($a=b=1$).}
\end{figure}

Figure~\ref{fig3} explores the variation of   several eigenenergy states and the g-factor vs. magnetic field for a symmetric quantum dot ($a=b=1$). In Fig.~\ref{fig3}(a), we plot several eigenenergy levels (ground, first, etc.)  vs. magnetic field for symmetric quantum dots with gate induced electric fields  $E=10^4$ V/cm (solid lines) and $E=5\times10^5$ V/cm (dashed lines). Here we find  level crossing shown by open circles at  magnetic field, $B=3.5$ T for the electric field $E=10^4$ V/cm (solid  lines) .  However, we find  an anticrossing  for  electric field $E=5\times10^5$ V/cm (dashed lines). The symbol $\times$ in Fig.~\ref{fig3} (a) represents the data from perturbation theory which is in agreement with the numerical simulations. In Fig.~\ref{fig3}(b), we plot the Land$\acute{\text{e}}$ g-factor vs. magnetic field at the electric fields $E=10^4$ V/cm (solid lines), $3\times10^5$ V/cm (dashed lines), $5\times10^5$ V/cm (dotted lines), $7\times10^5$ V/cm (dashed-dotted lines) and $10^6$ V/cm (dashed-dotted-dotted lines) for a symmetric quantum dot. It is clear  that there is a level crossing between the states $|0,0,-1/2>$ and  $|0,-1,+1/2>$ in the manipulation of the g-factor  at $B=3.5$ T for $E=10^4$ V/cm (solid lines). However, with the increase in the electric field, we have an anticrossing in the variation of the g-factor. This is also consistent with previously published work by R. de Sousa and S. Das Sarma~\cite{sousa03} and we consider this result as a benchmark  for our computational work.

\begin{figure*}
\includegraphics[width=19cm,height=6cm]{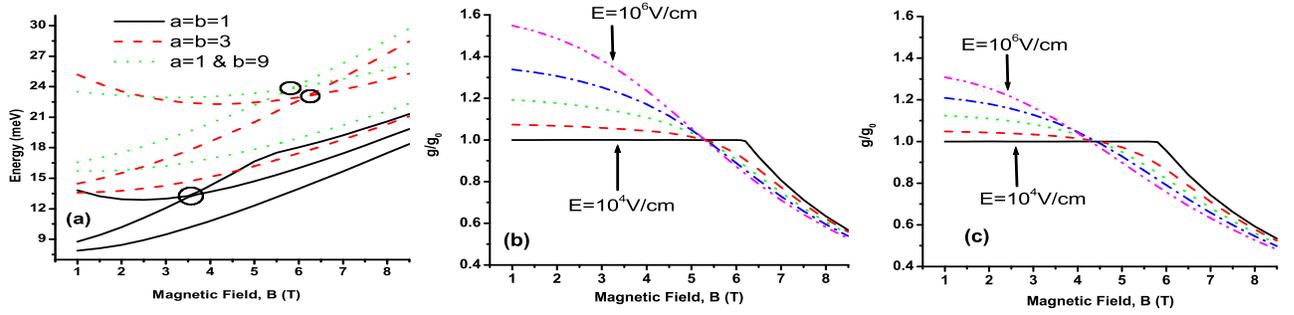}
\caption{\label{fig4} (Color online) (a) Several eigenenergy levels vs.  magnetic field at the  electric field $E=10^4$ V/cm  for both symmetric  ($a=b=1$ (solid lines), $a=b=3$ (dashed lines))  and asymmetric quantum dots ($a=1, b=9$, (dotted lines)). In (b) and  (c), we plot effective Land$\acute{\text{e}}$ g-factor vs.  magnetic fields. The electric fields are chosen as same as in Fig.~\ref{fig3}(b).  Also, in  (b) and (c),  we choose $a=b=3$ and $a=1, b=9$  respectively. In all cases,  we choose the quantum dot radius, $\ell_0=20$ nm.  }
\end{figure*}

In Fig.~\ref{fig4}, we analyze  anisotropy effects on the variation of the eigenenergy and of the Land$\acute{\text{e}}$ g-factor vs.  magnetic fields. In Fig.~\ref{fig4}(a), we plot the eigenenergy vs. magnetic field for  quantum dots in the potentials characterized by   $a=b=1$ (solid lines), $a=b=3$ (dashed lines) and  $a=1, b=9$ (dotted lines).  Here, we choose $E=10^4$ V/cm. We find a level crossing at $B=3.5$ T for the symmetric quantum dot ($a=b=1$). However, the level crossing extends approximately to  $B=6.2$ T for the symmetric quantum dots ($a=b=3$). The extension of the level crossing  to the larger magnetic fields is mainly due to the increase on the lateral size of the quantum dots. We quantify that the area of the quantum dots with ($a=b=3$) is $9$ times  larger than the dots with ($a=b=1$). For anisotropic quantum dots with the potential ($a=1, b=9$), we find the level crossing at around $B=6$ T. Note that the area of the asymmetric dots in the potential ($a=1, b=9$) is exactly equal to the area of the symmetric dots with the potential ($a=b=3$).
In Fig.~\ref{fig4}(b) and (c), we plot the variation in the Land$\acute{\text{e}}$ g-factor vs. magnetic field for  quantum dots in the confining potentials of $a=b=3$ and $a=1, b=9$. In both cases, we choose the electric fields as same as in Fig.~\ref{fig3}(b).
In Fig.~\ref{fig4} (b) for symmetric quantum dots (a=b=3), we find a crossover point (all the curves collapse to a single point) at  around $B=5$ T. However, this point decreases to approximately $B=4.5$ T for asymmetric quantum dots (a=1, b=9) as shown in Fig.~\ref{fig4}(c). At this point, the value of g-factor is equal to the bulk g-value of InAs dots  (i.e., $g=g_0=-15$) for all gate induced electric fields. Also, in Fig.~\ref{fig4}(c), we see that the anisotropic potential gives a quenching effect in the orbital angular momentum that reduces the electric field tunability of the g-factor.

\begin{figure*}
\includegraphics[width=19cm,height=6cm]{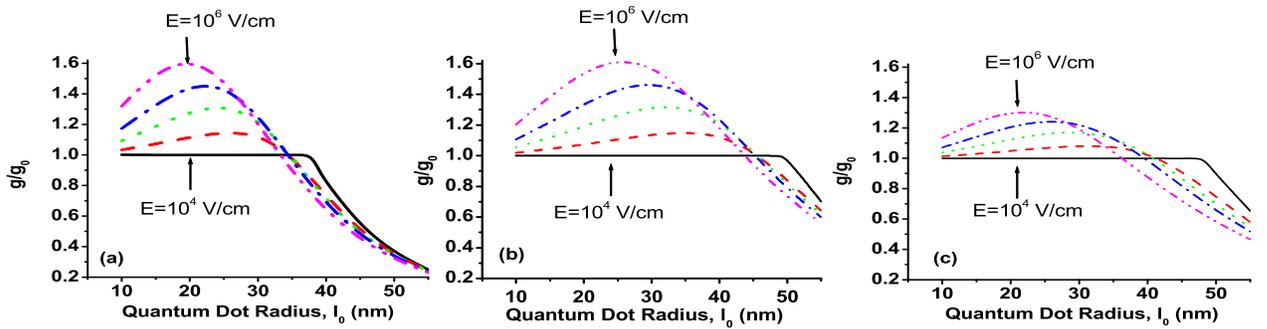}
\caption{\label{fig5} (color online) Land$\acute{\text{e}}$ g-factor  vs. quantum dot radius  for both symmetric ($a=b$)  and asymmetric ($a \neq b$)  quantum dots. Again, the electric fields are chosen as same as in Fig.~\ref{fig3}(b). Also we choose $a=b=1$, $a=b=3$ and  $a=1, b=9$ in (a),(b) and (c)  respectively. In all cases, we consider the magnetic field $B=1$ T.
}
\end{figure*}

Finally, Fig.~\ref{fig5} illustrates the manipulation of the Land$\acute{\text{e}}$ g-factor vs. quantum dot radii for both symmetric and asymmetric quantum dots. Again, the electric fields are chosen as same as in Fig.~\ref{fig3}(b).
In Fig.~\ref{fig5}(a) and (b), we consider isotropic confining potentials: $a=b=1$ and $a=b=3$. In Fig.~\ref{fig5}(b), level crossing point extends to the larger quantum dot radii compared to that of Fig.~\ref{fig5}(a) mainly due to the increase in the lateral size of the quantum dots. In Fig.~\ref{fig5}(c), we capture the anisotropic effect of the E-field tunability of the g-factor vs. quantum dot radii with the potential $a=1, b=9$. Again, we see that the anisotropic potential $a=1, b=9$ (Fig.~\ref{fig5}(c)) reduces the E-field tunability compared to that of isotropic potential $a=b=3$.
\begin{figure*}
\includegraphics[width=18cm,height=5cm]{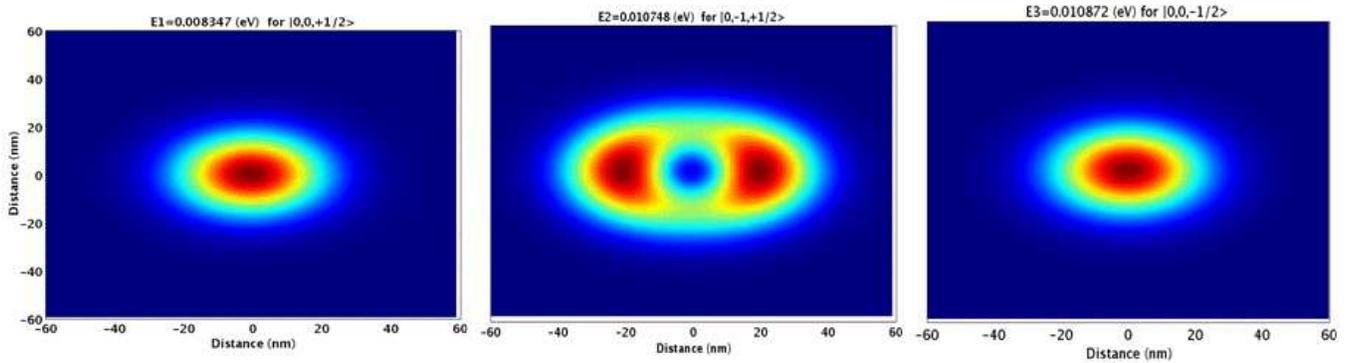}
\caption{\label{fig6} (color online) In-plane wavefunctions  in an asymmetric quantum dot. Here we choose $a=1.5$, $b=4$, $E=1.6\times 10^4$ V/cm, $B=2.9$ T and $\ell_0=28$ nm. These parameters  mimic the  wavefunctions of the quantum  states $|0,0,+1/2>~, |0,-1,+1/2>$ and $|0,0,-1/2>$ (from left to right) in Ref.~\onlinecite{takahashi10} of Fig.(2a). }
\end{figure*}
\begin{figure}
\includegraphics[width=9cm,height=6cm]{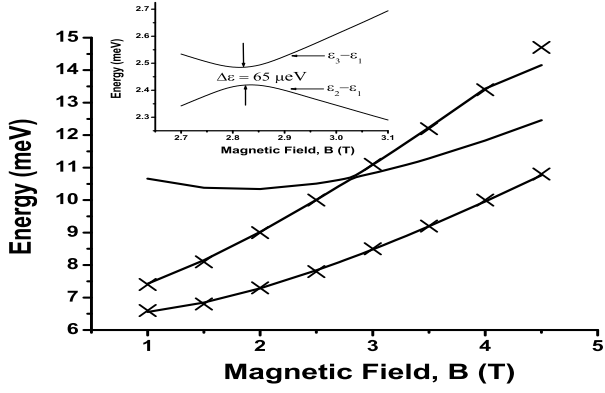}
\caption{\label{fig7} Eigenenergy of electron  states  $|0,0,+1/2>, |0,-1,+1/2>$ and $|0,0,-1/2>$ vs. magnetic field. The symbol $\times$ represents the data from perturbation theory. The inset shows the energy difference vs. magnetic field near the level crossing point. Here we estimate the size of the anticrossing point to be approximately 65 $\mu eV$. Here we again consider $a=1.5$, $b=4$, $E=1.6\times 10^4$ V/cm, and $\ell_0=28$ nm.
}
\end{figure}
\begin{figure}
\includegraphics[width=9.5cm,height=9cm]{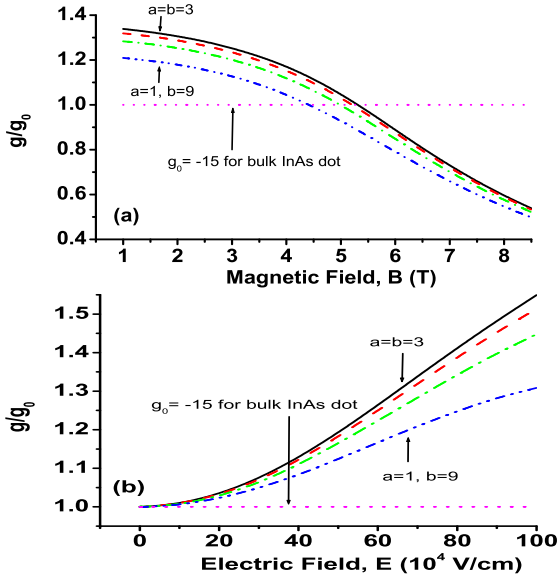}
\caption{\label{fig8} (Color online) Manipulation of the Land$\acute{\text{e}}$ g-factor vs. magnetic and electric fields for  quantum dots in the potential characterized by (top to bottom) $a=b=3$ (solid lines), $a=2, b=4.5$ (dashed lines), $a=1.5, b=6$ (dashed-dotted line) and $a=1, b=9$ (dashed-dotted-dotted lines).  We consider electric field, $E=7\times 10^5$ V/cm in case (a) and magnetic fields $B=1$ T in case (b). In both cases, we consider the quantum dot radius $\ell_0=20$ nm.
}
\end{figure}

Now we consider the wavefunctions  and estimate the size of the anticrossing point in an experimentally reported self assembled InAs quantum dot in Ref. \onlinecite{takahashi10}. The authors in  Ref. \onlinecite{takahashi10} characterize the lateral size of the quantum dots by two anisotropic gate potentials, $\hbar \omega_x=1.5$ meV and $\hbar \omega_y=4$ meV. This anisotropic potentials give $46$ nm lateral size of the quantum dots along the x-axis and $28$ nm along the y-axis in the plane of 2DEG. By substituting $a=1.5$, $b=4$ and $\ell_0=28$ nm in our theoretical model, this will  mimic the experimentally reported values of the lateral size of the quantum dots.
The authors in  Ref.~\onlinecite{takahashi10} also report that the height of the pyramidal shape quantum dots is $20$ nm. In our theoretical model, the quantum dot is formed in the plane of 2DEG so we choose the average height of the quantum dots as $10$ nm. In very crude approximation, this thickness of the 2DEG  is their  height of the pyramidal shape quantum dots.~\cite{williamson00}  Now, the  expression  $(2meE/\hbar^2)^{-1/3}$ in our theoretical model gives an estimate of the thickness of  2DEG which basically defines the size of the quantum dot along z-direction.
$10$ nm  vertical size of the quantum dot in the above expression gives an estimate of the electric fields, $E=1.6\times 10^4$ V/cm. We  find the anticrossing point at around $B=2.9$ Tesla and consider this value to capture the realistic wavefunctions  for the states  $|0,0,+1/2>, |0,0,-1/2>$ and $|0,-1,1/2>$. From Fig. \ref{fig6}, the first excited state wavefunction is  $|0,-1,+1/2>$ which is a clear indication of the level crossing point. In Fig. \ref{fig7}, we plot eigenenergy of the electron states $|0,0,+1/2>, |0,-1,+1/2>$ and $|0,0,-1/2>$ vs.  magnetic field. Again, the symbol $\times$ in Fig.~\ref{fig7} represents the data from perturbation theory for asymmetric potentials  which is in agreement with the numerical simulations. It can be seen that there is an anticrossing point at the magnetic field of around $2.9$ Tesla. Theoretically investigated level crossing point (B=2.9 T) is slightly smaller than the experimentally reported values (B=3.5 T, see Ref.~\onlinecite{takahashi10}) which indicates that the characterization of the lateral size of the quantum dots by anisotropic potentials has an estimate error of $20\%$.
The inset plot shows the magnified image near the level crossing point. Here we estimate the size of the anticrossing to be nearly $65$ $\mu$eV which is in agreement with the experimentally reported values in Ref.~\onlinecite{takahashi10}.

In Fig.~\ref{fig8}, we summarize our results for the manipulation of the Land$\acute{\text{e}}$ g-factor with respect to both magnetic and electric fields in isotropic and anisotropic quantum dots. In Fig.~\ref{fig8}(a), we plot the g-factor vs. magnetic field  at $E=7\times 10^5$ V/cm. Also, we plot the g-factor vs. electric field at $B=1$ T in Fig.~\ref{fig8}(b). In both cases, we compare the result for isotropic quantum dots ($a=b=3$, solid lines) and for anisotropic quantum dots $a=2, b=4.5$ (dashed lines), $a=1.5, b=6$ (dashed-dotted lines) and $a=1, b=9$ (dashed-dotted-dotted lines). Note that the quantum dots in the above confining potentials have the same area. At large magnetic fields,  $B\sim 5$ T in Fig.~\ref{fig8}(a) and small electric fields, $E\sim10^5$ V/cm in Fig.~\ref{fig8}(b), the effect of the resulting spin and orbital angular momentum of the quantum states on the eigenvalues   ($\varepsilon_{0,0,+1/2}-\varepsilon_{0,0,-1/2}$) are small, leading towards the bulk g-factor $i.e.$ $g/g_0\sim 1$ or $g=-15$. However, at small magnetic fields, $B\sim 1$ T in Fig.~\ref{fig8}(a) and large electric fields, $E\sim10^6$ V/cm in Fig.~\ref{fig8}(b), the effect on ($\varepsilon_{0,0,+1/2}-\varepsilon_{0,0,-1/2}$) is large enough to drag the g-factor towards the free electron value $i.e.$ $g/g_0\sim 2$.
Also, in both cases, if we compare the g-factor in both isotropic and anisotropic quantum dots, we find that anisotropic potentials ($a\neq b$) give the quenching effect in the orbital angular momentum  that  reduces  the variation in the B-field tunability and E-field tunability of the  Land$\acute{\text{e}}$ g-factor. This is also expected from Eqs.~\ref{g-anisotropic} and~\ref{g-isotropic}. For example, suppose  $B=1$ T, $\ell_0=20$ nm and $E=3\times10^5$ V/cm. Then from Eq.~\ref{g-anisotropic}, we find $g_{asym}=1.04$ for anisotropic quantum dots ($a=1, b=9$). However, from Eq.~\ref{g-isotropic}, we find $g_{sym}=1.07$ for isotropic quantum dots ( $a=b=3$). Also in Fig.~\ref{fig8} (a) and (b), it can be seen that $g_{sym}>g_{asym}$. Again, this quantitative analysis is in excellent agreement with our numerical simulations.

\section{Conclusions}
By utilizing both  analytical expressions and numerical simulations (based on Finite Element Method), from Figs.~\ref{fig1}-\ref{fig5},  we have shown that the Rashba spin-orbit coupling has the major contribution on the variation of the g-factor with electric fields  before  the level crossing or anticrossing occurs.
In Fig. \ref{fig6}, we have demonstrated the results of modeling of the realistic wavefunctions of the  states $|0,0,+1/2>, |0,-1,+1/2>$ and $|0,0,-1/2>$ near the level crossing point and estimate $65$ $\mu$eV as the size of the anticrossing in Fig. \ref{fig7}.
Finally, in Fig. \ref{fig8}, we have shown that  the anisotropic gate potential gives the quenching effect in the orbital angular momentum that reduces  the variation in the B-field tunability and E-field tunability of the g-factor.

\begin{acknowledgments}
 This work was supported by NSERC and CRC program, Canada.
\end{acknowledgments}

\appendix
\section{Energy spectrum of asymmetric quantum dot}
The Hamiltonian of an electron  in a quantum dot  in the plane of a $2$DEG  can be
written as
\begin{equation}
H_{xy} = \frac{\vec{P}^2_x+\vec{P}^2_y}{2 m} + {\frac{1}{2}} m \omega_o^2 (a x^2 + b y^2).
\label{hxy-a}
\end{equation}
The Eq.~(\ref{hxy-a}) can be exactly diagonalized~\cite{schuh85} by making a canonical transformation of position and momentum operators as
\begin{eqnarray}
x_1=\sqrt{\frac{\Omega_1}{\omega_0}}~x,\quad P_1=\sqrt{\frac{\omega_0}{\Omega_1}}P_x, \label{x-1}\\
x_2=\sqrt{\frac{\Omega_2}{\omega_0}}~y,\quad P_2=\sqrt{\frac{\omega_0}{\Omega_2}}P_y, \label{x-2}
\end{eqnarray}
where $\Omega_1=\omega_0\sqrt{a}$ and $\Omega_2=\omega_0\sqrt{b}$. Also, the Gauge potential can be written as
\begin{equation}
A_x=-\frac{x_2B\sqrt{\Omega_2\omega_0}}{\Omega_1+\Omega_2},~~~A_y=\frac{x_1B\sqrt{\Omega_1\omega_0}}{\Omega_1+\Omega_2}.\label{A-x}
\end{equation}
Substituting Eqs. (\ref{x-1},~\ref{x-2},~\ref{A-x}) into Eq.~(\ref{hxy-a}), we get the Hamiltonian in the form of:
\begin{equation}
H_{xy} = \frac{\Omega_1}{2m\omega_0}\left[P_1^2 + x_1^2 + e\left(P_2^2+x_2^2\right)+c\left(x_1P_2-x_2P_1\right)\right].
\label{hxy-b}
\end{equation}
The abbreviations used in Eq.~(\ref{hxy-b}) are as follows: $e=\Omega_2/\Omega_1$, $c=\left(2\omega_c\sqrt{\Omega_2/\Omega_1}\right)/\left[\omega_c^2+\left(\Omega_1+\Omega_2\right)^2\right]^{1/2}$. Also we use the relation $m\omega_0\gamma=1$, where $\gamma^2=1+\left[\omega_c^2/\left(\Omega_1+\Omega_2\right)^2\right]$.

The energy  spectrum of  Hamiltonian (\ref{hxy-b}) can be found as follows.
First, we need  to find the canonical transformation $U$ of the four-dimensional phase space, $P^t\equiv\left(P_x,P_y,x,y\right)$   which diagonalizes  the
quadratic form of the Hamiltonian (\ref{hxy-b}). To be more specific,  Hamiltonian (\ref{hxy-a})  can be written as
\begin{equation}
H_{xy}=\left(\frac{\Omega_1}{2m\omega_0}\right)P^tMP,~
\mathbf{M}=
\left(\begin{array}{cccc}
1 & 0 & 0 & -c/2 \\
0 & e & c/2 & 0 \\
0 & c/2 & 1 & 0 \\
-c/2 & 0 & 0 & e \\
\end{array}\right),
\label{hxy-c}
\end{equation}
where $t$ represents  the transpose of a vector. The orthogonal unitary matrix $U$ which exactly diagonalizes the matrix M can be written as,
\begin{equation}
\mathbf{U}=\frac{1}{\left(s_+-s_-\right)}
\left(\begin{array}{cccc}
1 & 1 & -s_- & -s_+ \\
1 & -1 & s_+ & -s_- \\
s_- & s_+ & 1 & 1 \\
-s_+ & s_- & 1 & -1 \\
\end{array}\right),
\label{U}
\end{equation}
where $cs_{\pm}\equiv e-1\pm d$ and
\begin{equation}
d=\sqrt{\frac{4\omega_c^2\Omega_2/\Omega_1}{\omega_c^2+\left(\Omega_1+\Omega_2\right)^2}+\left(1-\frac{\Omega_2}{\Omega_1}\right)^2}.
\label{d}
\end{equation}
Also, the expression for $s_{\pm}$ is written in Eq. (\ref{s-pm}). In terms of rotated operators   $P'=UP$, the Hamiltonian (\ref{hxy-c}) can be written as
\begin{eqnarray}
H_{xy}&=&\left(\frac{\Omega_1}{2m\omega_0}\right)\nonumber\\
&&\left[\frac{1}{2}(cs_-+2)\left(p_x^{'2}+x^{'2}\right)+\frac{1}{2}(cs_++2)\left(p_y^{'2}+y^{'2}\right)\right].\nonumber
\label{hxy-d}
\end{eqnarray}
The above Hamiltonian is identified as the  superposition of two independent harmonic oscillators and its energy spectrum can be written as
\begin{equation}
\varepsilon_{n_+n_-}=\left(n_++n_-+1\right)\hbar\omega_++\left(n_+-n_-\right)\hbar\omega_-, \label{epsilon-a}\\
\end{equation}
provided that $\left[a_{\pm}^\dagger,a_{\pm}\right]=1$. Also $\omega_{\pm}=\frac{1}{2}\left[\omega_c^2+\left(\Omega_1\pm\Omega_2\right)^2\right]^{1/2}$.

\end{document}